\documentstyle[12pt,amssymb,amsmath]{article}
\newtheorem{theorem}{Theorem}[section]
\newtheorem{lemma}{Lemma}[section]

\newcommand{\p}{\partial}

\newcommand{\s}{\sum_{i=1}^N}

\begin{document}

\title{ {\bf B\"{a}cklund transformations for the constrained dispersionless hierarchies and dispersionless hierarchies with
self-consistent sources}  }

\author{ {\bf  Ting Xiao\dag
    \hspace{1cm} Yunbo Zeng\dag\dag } \\
    {\small {\it
    Department of Mathematical Sciences, Tsinghua University,
    Beijing 100084, China}} \\
    {\small {\it \dag
     Email: xiaoting97@mails.tsinghua.edu.cn}}\\
    {\small {\it \dag\dag
     Email: yzeng@math.tsinghua.edu.cn}} }

\date{}
\maketitle
\renewcommand{\theequation}{\arabic{section}.\arabic{equation}}

\begin{abstract}
The B\"{a}cklund transformations between the constrained
dispersionless KP hierarchy (cdKPH) and the constrained
dispersionless mKP hieararchy (cdmKPH) and between the
dispersionless KP hieararchy with self-consistent sources (dKPHSCS)
and the dispersionless mKP hieararchy with self-consistent sources
(dmKPHSCS) are constructed. The auto-B\"{a}cklund transformations
for the cdmKPH and for the dmKPHSCS are also formulated.
\end{abstract}

\hskip\parindent


\section{Introduction}
\setcounter{equation}{0} \hskip\parindent The study of constrained
integrable systems has attracted much attention in the last two
decades (see [1-5] and the references therein). It was shown that
some important integrable hierarchies of (1+1)-dimensions are
induced by constraints of integrable hierarchies of
(2+1)-dimensions. For example, the famous AKNS hierarchy,
Yajima-Oikawa (Y-O) hierarchy are results of constraints to the KP
hierarchy \cite{Oevel933,Cheng95} while the Kaup-Broer hierarchy is
a result of constraint to the mKP hierarchy \cite{Oevel933}. Some
integrable properties of the constrained hierarchies, such as the
Lax representations, bi-Hamiltonian structures etc. can be inherited
from those of the original hierarchies. So the constraint offers us
a bridge connecting integrable hierarchies of higher dimensions to
those of lower dimensions.

The soliton equations with self-consistent sources (SESCS) are
important integrable models in many fields of physics, such as
hydrodynamics, state physics, plasma physics, etc (see [6-10] and
the references therein). For example, the KP equation with
self-consistent sources (KPESCS) describes the interaction of a long
wave with a short-wave packet propagating on the x,y plane at an
angle to each other \cite{Mel'nikov87}. In our general approach
proposed recently in [11-17], the constrained integrable hierarchy
may be regarded as the stationary system of the corresponding
integrable hierarchy with self-consistent sources. From this
observation, the integrable hierarchies with self-consistent sources
and their Lax representations can be constructed from the
corresponding constrained integrable hierarchies. The KP hierarchy
with self-consistent sources (KPHSCS) and the modified KP hierarchy
with self-consistent sources (mKPHSCS) are constructed in this way
and their interesting solutions are obtained via generalized Darboux
transformation \cite{Xiao20041,Xiao20042}. In this sense, the
soliton hierarchies with self-consistent sources may be viewed as
integrable generalizations of the original soliton hierarchies.

B\"{a}cklund transformation is a powerful tool to investigate the
soliton hierarchies. The (2+1) dimensional B\"{a}cklund
transformations employed were constructed by the so-called Dressing
Method based on appropriate gauge transformation \cite{Boiti85}.
This dressing approach had been earlier used to construct
auto-B\"{a}cklund transformations for the KP, two-dimensional three
wave and Davey-Stewartson equations in turn \cite{Levi81}.
Konopelchenko, Oevel et al extended this method to investigate some
important 2+1 dimensional integrable hierarchies constructed in the
framework of Sato theory, such as the KP hierarchy, the mKP
hierarchy and the Dym hierarchy [20-25]. Also, gauge transformations
are applicable to investigate the constrained KP and mKP hierarchies
\cite{Oevel933,Oevel931,Shaw971,Shaw972}. Lately, utilizing
eigenfunctions and adjoint eigenfunctions, we developed this idea to
construct several types of auto-B\"{a}cklund transformations for the
KPHSCS and mKPHSCS respectively. The B\"{a}cklund transformations
from the KPHSCS to mKPHSCS are also constructed in this way
\cite{Xiao20053}.

Recently, we proposed a new method to construct the constrained
dispersionless hierarchies by means of the quasiclassical limit of
the squared eigenfunction symmetry constraint of the corresponding
dispersionfull hierarchies \cite{Xiao20051,Xiao20052}. The
constrained dispersionless KP hierarchy (cdKPH) and mKP hierarchy
(cdmKPH) are constructed in this way in \cite{Xiao20051,Xiao20052}
and we find that the original dispersionless hierarchies of
(2+1)-dimensions are decomposed into two commutative
(1+1)-dimensional hierarchies of hydrodynamical type which offers us
a new approach to study the dispersionless hierarchies of
(2+1)-dimensions. The corresponding dispersionless hierarchies with
self-consistent sources together with their associated conservation
equations are also obtained therefore which can be solved by the
hodograph reduction method. Based on these facts, the dispersionless
hierarchies with self-consistent sources may be viewed as integrable
generalizations of the original dispersionless hierarchies. In
contrast with the dispersionfull case, the connections between these
constrained dispersionless hierarchies and dispersionless
hierarchies with self-consistent sources have not been studied yet.
It is attractive for us to reveal the hidden relations between these
dispersionless hierarchies. Compared to the dressing operation to
the Sato operators for the dispersionfull hierarchies, a 'dressing'
operation to the Sato functions for the dispersionless hierarchies
\cite{Takasaki1995,Chang2000} is used here. Motivated by the
B\"{a}cklund transformation between dKP and dmKP hierarchies
proposed by Chang and Tu \cite{Chang2000}, we succeed to construct
the B\"{a}cklund transformations between cdKPH and cdmKPH and
between dKPHSCS and dmKPHSCS respectively. The auto-B\"{a}cklund
transformations for cdmKPH and for dmKPHSCS are also constructed.
The B\"{a}cklund transformations between relevant dispersionless
hierarchies reveal the intimate relations between them and the
auto-B\"{a}cklund transformations for cdmKPH and dmKPHSCS reveal the
'gauge' invariance of them.

The paper will be organized as follows. In section 2, we review the
definitions of the cdKPH, dKPHSCS and cdmKPH, dKPHSCS respectively.
In section 3, the B\"{a}cklund transformations from cdKPH to cdmKPH
and from dKPHSCS to dmKPHSCS are constructed. The auto-B\"{a}cklund
transformations for cdmKPH and dmKPHSCS are formulated respectively
in Section 4.

\section{The constrained dispersionless hierarchies and dispersionless hierarchies with self-consistent sources}
\setcounter{equation}{0} \hskip\parindent
1. cdKPH and dKPHSCS.\\
The dispersionless KP hierarchy is defined as
\begin{equation}
\label{17}
    \partial_{T_m}{\mathcal{K}} = \{{\mathcal{B}}_m, {\mathcal{K}}\},\ \ \ {\mathcal{B}}_m=({\mathcal{K}}^m)_{\geq
    0},
\end{equation}
where the Sato function ${\mathcal{K}}$ is given by
\begin{equation}
\label{18}
    {\mathcal{K}} = p+\sum_{i=1}^\infty
    U_i(T)p^{-i},
\end{equation}
with $T=(T_1=X,T_2,\cdots)$ and the Poisson bracket is defined as
\begin{equation}
\label{181} \{A(p,x), B(p,x)\} = \frac{\partial A}{\partial p}
\frac{\partial B}{\partial x}-\frac{\partial A}{\partial
x}\frac{\partial B}{\partial p}.
\end{equation}
The Sato function ${\mathcal{K}}$ (\ref{18}) has a type of
constraint as
\begin{equation}
\label{19}
{\mathcal{K}}^n={\mathcal{B}}_n+\sum_{i=1}^N\frac{\alpha_i}{p-\beta_i},\
\ {\mathcal{B}}_n=({\mathcal{K}}^n)_{\geq 0},\ \ n\in \mathbb{N},
\end{equation}
where $\alpha_i$, $\beta_i$ satisfy
\begin{subequations}
\begin{equation}
     \alpha_{i,T_m}=[\alpha_i(\frac{\partial}{\partial p}{\mathcal{B}}_m(p))|_{p=\beta_i}]_X,
\end{equation}
\begin{equation}
    \beta_{i,T_m}=[{\mathcal{B}}_m(p)|_{p=\beta_i}]_X,\ \ \ \
    i=1,...,N, m\in\mathbb{N},
\end{equation}
\end{subequations}
with ${\mathcal{B}}_m=[({\mathcal{K}}^n)^{\frac{m}{n}}]_{\geq 0}$.
It is shown in \cite{Xiao20051} that the constrained Sato function
${\mathcal{K}}^n$ (\ref{19}) can be obtained by the quasiclassical
limit to the constrained Sato operator for the KP hierarchy.\\
The constrained disperionless KP
hierarchy (cdKPH) is \cite{Xiao20051}
\begin{subequations}
\label{110}
\begin{equation}
\label{1101}
     ({\mathcal{K}}^n)_{T_m} =
     \{{\mathcal{B}}_m,{\mathcal{K}}^n\},
\end{equation}
\begin{equation}
\label{1102}
     \alpha_{i,T_m}=[\alpha_i(\frac{\partial}{\partial p}{\mathcal{B}}_m(p))|_{p=\beta_i}]_X,
\end{equation}
\begin{equation}
\label{1103}
    \beta_{i,T_m}=[{\mathcal{B}}_m(p)|_{p=\beta_i}]_X,\ \ \ \
    i=1,...,N, m\in\mathbb{N}.
\end{equation}
\end{subequations}
Requiring $m<n$, the dKP hierarchy with self-consistent sources
(dKPHSCS) is defined by \cite{Xiao20051}
\begin{subequations}
\label{111}
\begin{equation}
\label{1111}
     ({\mathcal{B}}_m)_{T_n}-({\mathcal{K}}^n)_{T_m}+ \{{\mathcal{B}}_m,{\mathcal{K}}^n\}=0,
\end{equation}
\begin{equation}
\label{1112}
     \alpha_{i,T_m}=[\alpha_i(\frac{\partial}{\partial p}{\mathcal{B}}_m(p))|_{p=\beta_i}]_X,
\end{equation}
\begin{equation}
\label{1113}
    \beta_{i,T_m}=[{\mathcal{B}}_m(p)|_{p=\beta_i}]_X,\ \
    i=1,...,N.
\end{equation}
\end{subequations}
The conservation equations (or equations of Hamilton-Jacobi type)
for the dKPHSCS (\ref{111}) are
\begin{subequations}
\label{112}
\begin{equation}
\label{1121}
     p_{T_m}=[{\mathcal{B}}_m(p)]_X,
\end{equation}
\begin{equation}
\label{1122}
     p_{T_n}=[{\mathcal{K}}^n(p)]_X=[{\mathcal{B}}_n(p)+\sum_{i=1}^N\frac{\alpha_i}{p-\beta_i}]_X,
\end{equation}
\end{subequations}
i.e., under (\ref{1112}) and (\ref{1113}), the compatibility of
(\ref{1121}) and (\ref{1122}) will give (\ref{1111}). For example,
when $m=2$, $n=3$, (\ref{19}) becomes
\begin{equation}
\nonumber
     {\mathcal{K}}^3=p^3+3U_1p+3U_2+\sum_{i=1}^N\frac{\alpha_i}{p-\beta_i},
\end{equation}
and (\ref{111}) turns into the dKP equation with self-consistent
sources (dKPESCS) ($U_2$ is eliminated by
$U_{2,X}=\frac{1}{2}U_{1,Y}$ and $Y=T_2$, $T=T_3$) \cite{Xiao20051}
\begin{subequations}
\label{113}
\begin{equation}
\label{1131}
     (U_{1,T}-3U_1U_{1,X}+\sum_{i=1}^N\alpha_{i,X})_X=\frac{3}{4}U_{1,YY},
\end{equation}
\begin{equation}
\label{1132} \alpha_{i,Y}=2(\alpha_i\beta_i)_X,
\end{equation}
\begin{equation}
\label{1133}
    \beta_{i,Y}=(\beta_i^2+2U_1)_X,\ \ i=1,...,N.
\end{equation}
\end{subequations}
2. The cmdKPH and dmKPHSCS.\\
Given the Sato function as
\begin{equation}
\label{210}
    {\mathcal{L}} = p+\sum_{i=0}^\infty
    V_i(T)p^{-i},
\end{equation}
the dispersionless mKP hierarchy is defined as
\begin{equation}
\label{211}
    \partial_{T_n}{\mathcal{L}} = \{{\mathcal{Q}}_n,
    {\mathcal{L}}\}.
\end{equation}
As shown in \cite{Xiao20052}, the quasiclassical limit to the
constrained Sato operator for the mKP hierarchy will give rise to
the constrained Sato function ${\mathcal{L}}^n$ for the
dispersionless mKP hierarchy as
\begin{equation}
\label{212}
{\mathcal{L}}^n={\mathcal{Q}}_n-\sum_{i=1}^N(\frac{a_i}{p_i}+\frac{a_i}{p-p_i}),\
\ {\mathcal{Q}}_n=({\mathcal{L}}^n)_{\geq 1},
\end{equation}
where $a_i$, $p_i$ satisfy
\begin{subequations}
\label{213}
\begin{equation}
\label{2131}
 a_{i,T_k}=[a_i(\frac{\partial}{\partial p}{\mathcal{Q}}_k(p))|_{p=p_i}]_X,
\end{equation}
\begin{equation}
\label{2132}
   p_{i,T_k}=[{\mathcal{Q}}_k(p)|_{p=p_i}]_X,\ \
    i=1,\cdots,N,
\end{equation}
\end{subequations}
with ${\mathcal{Q}}_k=[({\mathcal{L}}^n)^{\frac{k}{n}}]_{\geq 1}$.
The constrained dispersionless mKP hierarchy (cdmKPH) is defined as
\cite{Xiao20052}
\begin{subequations}
\label{214}
\begin{equation}
\label{2141}
     ({\mathcal{L}}^n)_{T_k} = \{{\mathcal{Q}}_k,{\mathcal{L}}^n\},
\end{equation}
\begin{equation}
\label{2142}
     a_{i,T_k}=[a_i(\frac{\partial}{\partial p}{\mathcal{Q}}_k(p))|_{p=p_i}]_X,
\end{equation}
\begin{equation}
\label{2143}
    p_{i,T_k}=[{\mathcal{Q}}_k(p)|_{p=p_i}]_X,\ \
    i=1,...,N.
\end{equation}
\end{subequations}
When adding the term $({\mathcal{Q}}_k)_{T_n}$ to the right hand
side of (\ref{2141}) and requiring $k<n$, we will obtain the dmKP
hierarchy with self-consistent sources (dmKPHSCS) as
\cite{Xiao20052}
\begin{subequations}
\label{215}
\begin{equation}
\label{2151}
     ({\mathcal{Q}}_k)_{T_n}-({\mathcal{L}}^n)_{T_k}+ \{{\mathcal{Q}}_k,{\mathcal{L}}^n\}=0,
\end{equation}
\begin{equation}
\label{2152}
     a_{i,T_k}=[a_i(\frac{\partial}{\partial p}{\mathcal{Q}}_k(p))|_{p=p_i}]_X,
\end{equation}
\begin{equation}
\label{2153}
    p_{i,T_k}=[{\mathcal{Q}}_k(p)|_{p=p_i}]_X,\ \
    i=1,...,N.
\end{equation}
\end{subequations}
It is not difficult to prove that under (\ref{2152}) and
(\ref{2153}), the equation (\ref{2151}) will be obtained by the
compatibility of the following conservation equations
\begin{subequations}
\label{216}
\begin{equation}
\label{2161}
     p_{T_k}=[{\mathcal{Q}}_k(p)]_X,
\end{equation}
\begin{equation}
\label{2162}
     p_{T_n}=[{\mathcal{L}}^n(p)]_X=[{\mathcal{Q}}_n(p)-\sum_{i=1}^N(\frac{a_i}{p_i}+\frac{a_i}{p-p_i})]_X.
\end{equation}
\end{subequations}
For example, when $k=2$, $n=3$,
\begin{equation}
\nonumber
     {\mathcal{L}}^3=p^3+3V_0p^2+(3V_0^2+V_1)p-\sum_{i=1}^N(\frac{a_i}{p_i}+\frac{a_i}{p-p_i}),
\end{equation}
(\ref{215}) becomes the dmKP equation with self-consistent sources
(dmKPESCS) ($Y=T_2$, $T=T_3$, $V=V_0$ and $V_1$ is eliminated by
$V_{1,X}=\frac{3}{2}V_{0,Y}-\frac{3}{2}(V_0^2)_X$) \cite{Xiao20052}
\begin{subequations}
\label{217}
\begin{equation}
\label{2171}
     2V_T-\frac{3}{2}D_X^{-1}(V_{YY})-3V_XD_X^{-1}(V_Y)+3V^2V_X-2\s(\frac{a_i}{p_i})_X=0,
\end{equation}
\begin{equation}
\label{2172}
     a_{i,Y}=2[a_i(p_i+V)]_X,
\end{equation}
\begin{equation}
\label{2173}
    p_{i,Y}=(p_i^2+2Vp_i)_X,\ \ i=1,...,N.
\end{equation}
\end{subequations}

\section{B\"{a}cklund transformations from cdKPH to cdmKPH and from dKPHSCS to dmKPHSCS} \setcounter{equation}{0} \hskip\parindent

Given $\phi(T)$, an arbitrary function of $T=(T_1=X,T_2,\cdots)$ and
a function $\mu(p,T)$ as
\begin{equation}
\label{31}
     \mu=p^l+a_{l-1}p^{l-1}+\cdots+a_0+\frac{a_{-1}}{p}+\frac{a_{-2}}{p^2}+\cdots,\
     \ a_i=a_i(T),\ \ i\leq l-1,
\end{equation}
we define \cite{Takasaki1995,Chang2000}
\begin{equation}
\label{32}
\tilde{\mu}=e^{-ad\phi}(\mu)=\mu-\{\phi,\mu\}+\frac{1}{2!}\{\phi,\{\phi,\mu\}\}-\frac{1}{3!}\{\phi,\{\phi,\{\phi,\mu\}\}\}+\cdots,
\end{equation}
where the Poisson bracket is defined by (\ref{181}). A simple
calculation shows
\begin{equation}
\label{33}
\tilde{\mu}=\sum_{n=0}^{\infty}\frac{1}{n!}(\phi_X)^n\p_p^n\mu.
\end{equation}

\begin{lemma}\cite{Chang2000}
\begin{subequations}
\label{34}
\begin{equation}
\label{341}
(i)e^{-ad\phi}(\mu(p,T)\nu(p,T))=e^{-ad\phi}(\mu(p,T))e^{-ad\phi}(\nu(p,T)),\
\ \ \ \ \ \ \ \ \ \ \ \ \ \ \ \ \ \ \ \ \ \ \ \ \
\end{equation}
\begin{equation}
\label{342} (ii)\tilde{\mu}_{\geq 1}=e^{-ad\phi}(\mu_{\geq
0})-\mu_{\geq 0}|_{p=\phi_X}=e^{-ad\phi}(\mu_{\geq 1})-\mu_{\geq
1}|_{p=\phi_X},\ \ \ \ \ \ \ \ \ \ \ \ \ \ \ \ \ \ \ \ \ \ \
\end{equation}
\begin{equation}
\label{343}
\begin{array}{lll}
(iii)\tilde{\mu}_{T_q}-\{(\tilde{\mu}^{\frac{q}{l}})_{\geq
1},\tilde{\mu}\}&=&e^{-ad\phi}(\mu_{T_q}-\{(\mu^{\frac{q}{l}})_{\geq
0},\mu\})-\{\phi_{T_q}-(\mu^{\frac{q}{l}})_{\geq
0}|_{p=\phi_X},\tilde{\mu}\}\\
&=&e^{-ad\phi}(\mu_{T_q}-\{(\mu^{\frac{q}{l}})_{\geq
1},\mu\})-\{\phi_{T_q}-(\mu^{\frac{q}{l}})_{\geq
1}|_{p=\phi_X},\tilde{\mu}\}.
\end{array}
\end{equation}
\end{subequations}
\end{lemma}
\begin{theorem}({\bf{B\"{a}cklund transformation from cdKPH to
cdmKPH}})\\
Let ${\mathcal{K}}^n$ of (\ref{19}) satisfy the cdKPH (\ref{110}).
$\phi$ is a function of $T$ satisfying
\begin{subequations}
\label{35}
\begin{equation}
\label{351} {\mathcal{B}}_n|_{p=\phi_X}+\s
\frac{\alpha_i}{\phi_X-\beta_i}=0,
\end{equation}
\begin{equation}
\label{352} \phi_{T_k}={\mathcal{B}}_k|_{p=\phi_X},\ \ \ \
k\in\mathbb{N}.
\end{equation}
\end{subequations}
Define
$$G_1[\phi]:{\mathcal{K}}^n\mapsto {\mathcal{L}}^n=e^{-ad\phi}({\mathcal{K}}^n),$$
then $${\mathcal{L}}^n={\mathcal{Q}}_n-\s
(\frac{a_i}{p_i}+\frac{a_i}{p-p_i}),\ \ \
{\mathcal{Q}}_n=({\mathcal{L}}^n)_{\geq 1},$$ where $a_i=-\alpha_i$,
$p_i=\beta_i-\phi_X$ will satisfy the cdmKPH (\ref{214}).
\end{theorem}
{\bf{Proof}}:
\begin{equation}
\begin{array}{lll}
({\mathcal{L}}^n)_{\leq 0}&=&[e^{-ad\phi}({\mathcal{K}}^n)]_{\leq
0}=e^{-ad\phi}(({\mathcal{K}}^n)_{<0})+[e^{-ad\phi}(({\mathcal{K}}^n)_{\geq 0})]_0\\
&=&e^{-ad\phi}(\s
\frac{\alpha_i}{p-\beta_i})+({\mathcal{K}}^n)_{\geq
0}|_{p=\phi_X}\\
&=&\s
\frac{\alpha_i}{p-(\beta_i-\phi_X)}+{\mathcal{B}}_n|_{p=\phi_X}\\
&=&-\s(\frac{a_i}{p_i}+\frac{a_i}{p-p_i}).
\end{array}
\end{equation}
Furthermore, we have
\begin{equation}
\label{36}
\begin{array}{lll}
{\mathcal{Q}}_k&=&[({\mathcal{L}}^n)^{\frac{k}{n}}]_{\geq
1}=\{[e^{-ad\phi}({\mathcal{K}}^n)]^{\frac{k}{n}}\}_{\geq
1}=[e^{-ad\phi}(({\mathcal{K}}^n)^{\frac{k}{n}})]_{\geq 1}\\
&=&e^{-ad\phi}({\mathcal{B}}_k)-{\mathcal{B}}_k|_{p=\phi_X}=\sum_{l=0}^{\infty}\frac{1}{l!}\phi_X^l\p_{p}^l({\mathcal{B}}_k)-{\mathcal{B}}_k|_{p=\phi_X},
\end{array}
\end{equation}
\begin{equation}
\label{37}
({\mathcal{Q}}_k)_p=[e^{-ad\phi}({\mathcal{B}}_{k})]_p=e^{-ad\phi}({\mathcal{B}}_{k,p})=\sum_{l=0}^{\infty}\frac{1}{l!}\phi_X^l\p_p^l({\mathcal{B}}_{k,p}).
\end{equation}
So
\begin{subequations}
\label{38}
\begin{equation}
\label{381}
{\mathcal{Q}}_k|_{p=p_i}={\mathcal{B}}_k|_{p=\beta_i}-{\mathcal{B}}_k|_{p=\phi_X},
\end{equation}
\begin{equation}
\label{382}
({\mathcal{Q}}_k)_p|_{p=p_i}=\sum_{l=0}^{\infty}\frac{1}{l!}\phi_X^l\p_p^l({\mathcal{B}}_{k,p})|_{p=p_i}={\mathcal{B}}_{k,p}|_{p=\beta_i}.
\end{equation}
\end{subequations}
From Lemma 4.1 and (\ref{38}), we can see that
\begin{subequations}
\label{39}
\begin{equation}
\label{391}
({\mathcal{L}}^n)_{T_k}-\{{\mathcal{Q}}_k,{\mathcal{L}}^n\}=e^{-ad\phi}[({\mathcal{K}}^n)_{T_k}-
\{{\mathcal{B}}_k,{\mathcal{K}}^n\}]-\{\phi_{T_k}-{\mathcal{B}}_k|_{p=\phi_X},{\mathcal{L}}^n\}=0,
\end{equation}
\begin{equation}
\label{392}
a_{i,T_k}=-\alpha_{i,T_k}=-(\alpha_i{\mathcal{B}}_{k,p}|_{p=\beta_i})_X=(a_i({\mathcal{Q}}_k)_p|_{p=p_i})_X,\
\ \ \ \ \ \ \ \ \ \ \ \ \ \ \ \ \ \ \ \ \ \ \ \ \ \ \ \ \
\end{equation}
\begin{equation}
\label{393}
p_{i,T_k}=\beta_{i,T_k}-(\phi_X)_{T_k}=({{\mathcal{B}}}_k|_{p=\beta_i}-\phi_{T_k})_X=({{\mathcal{B}}}_k|_{p=\beta_i}-{{\mathcal{B}}}_k|_{p=\phi_X})_X=({\mathcal{Q}}_k|_{p=p_i})_X,
\end{equation}
\end{subequations}
i.e., ${\mathcal{L}}^n$, $a_i$, $p_i$ satisfy the cdmKPH
(\ref{214}).\\
This completes the proof.

\begin{theorem}({\bf{B\"{a}cklund transformation from dKPHSCS to
dmKPHSCS}})\\
Let ${\mathcal{K}}^n$ of (\ref{19}) satisfy the dKPHSCS (\ref{111})
and $p$ satisfy the corresponding conservation equations
(\ref{112}). $\phi$ is a function of $T$ satisfying
\begin{subequations}
\label{310}
\begin{equation}
\label{3101} \phi_{T_k}={\mathcal{B}}_k|_{p=\phi_X},
\end{equation}
\begin{equation}
\label{3102} \phi_{T_n}={\mathcal{B}}_n|_{p=\phi_X}+\s
\frac{\alpha_i}{\phi_X-\beta_i}.
\end{equation}
\end{subequations}
Define \begin{equation} \label{3103}
G_2[\phi]:{\mathcal{K}}^n\mapsto
{\mathcal{L}}^n=[e^{-ad\phi}({\mathcal{K}}^n)]_{\geq 1}-\s
(\frac{a_i}{p_i}+\frac{a_i}{p-p_i}),
\end{equation}
where $a_i=-\alpha_i$, $p_i=\beta_i-\phi_X$, then ${\mathcal{L}}^n$
will satisfy the dmKPHSCS (\ref{215}).
\end{theorem}
{\bf{Proof}}: Define $\tilde{p}=p-\phi_X$. For $k<n$,
\begin{equation}
\label{311}
\begin{array}{lll}
{\mathcal{Q}}_k&=&[({\mathcal{L}}^n)^{\frac{k}{n}}]_{\geq
1}=[(({\mathcal{L}}^n)_{\geq 1})^{\frac{k}{n}}]_{\geq 1}=
\{[(e^{-ad\phi}({\mathcal{K}}^n))_{\geq 1}]^{\frac{k}{n}}\}_{\geq
1}=[(e^{-ad\phi}({\mathcal{K}}^n))^{\frac{k}{n}}]_{\geq 1}\\
&=&[e^{-ad\phi}(({\mathcal{K}}^n)^{\frac{k}{n}})]_{\geq
1}=e^{-ad\phi}({\mathcal{B}}_k)-{\mathcal{B}}_k|_{p=\phi_X}.
\end{array}
\end{equation}
So as proved in Theorem 4.1, $a_i$, $p_i$, $i=1,\cdots,N$ will
satisfy (\ref{2152}) and (\ref{2153}) respectively. Similarly we can
prove $\tilde{p}$ satisfies (\ref{2161}).
\begin{equation}
\begin{array}{lll}
\tilde{p}_{T_n}&=&p_{T_n}-\phi_{X,T_n}=({\mathcal{B}}_n(p)+\s\frac{\alpha_i}{p-\beta_i})_X-({\mathcal{B}}_n|_{p=\phi_X}+\s\frac{\alpha_i}{\phi_X-\beta_i})_X\\
&=&({\mathcal{B}}_n(p)-{\mathcal{B}}_n|_{p=\phi_X})_X+(\s\frac{\alpha_i}{p-\beta_i}+\s\frac{\alpha_i}{p_i})_X\\
&=&[{\mathcal{Q}}_n(p)|_{p=\tilde{p}}]_X-\s(\frac{a_i}{\tilde{p}-p_i}+\frac{a_i}{p_i})_X,
\end{array}
\end{equation}
so $\tilde{p}$ satisfies (\ref{2162}).\\
This completes the proof.\\
\\
{\bf{Example 1}:(B\"{a}cklund transformation from dKPESCS to
dmKPESCS)}\\
When $n=3$, $G_2[\phi]$ will offers us a B\"{a}cklund transformation
from the dKPESCS (\ref{113}) to the dmKPESCS
(\ref{217}). From (\ref{3103}), $V=V_0=\phi_X$.\\
{\bf{(1)}} If we take a simple solution of the dKPESCS (\ref{113})
as $U_1=0$, $\alpha_i=\beta_i=0$, $i=1,\cdots,N$, then $\phi(X,Y,T)$
correspondingly has to satisfy
\begin{equation}
\label{3111} \phi_Y=\phi_X^2,\ \ \ \ \ \ \ \phi_T=\phi_X^3.
\end{equation}
 Define
$q=\phi_X$, then $q$ satisfies the following equations of
hydrodynamical type
\begin{equation}
\label{312} q_Y=2qq_X,\ \ \ \ \ \ \ \ q_T=3q^2q_X.
\end{equation}
The solution of (\ref{312}) can be determined by the hodograph
equation
\begin{equation}
\label{313} F(q)=X+2qY+3q^2T,
\end{equation}
where $F(q)$ is an arbitrary function of $q$. When $F(q)=0$,
$q=\frac{-Y\pm \sqrt{Y^2-3TX}}{3T}$, then $V=\phi_X=q$,
$a_i=-\alpha_i=0$, $p_i=\beta_i-\phi_X=-q$, $i=1,\cdots,N$ is a
degenerate solution of the dmKPESCS (\ref{217}) ('degenerate' means
since $a_i=0$, (\ref{2171}) degenerates to the dmKP
equation).\\
{\bf{(2)}} If we choose $U_1=0$, $\alpha_i=1$, $\beta_i=0$,
$i=1,\cdots,N$, then the hodograph equation (\ref{313})
correspondingly changes to
\begin{equation}
\label{314} F(q)=X+2qY+(3q^2-\frac{N}{q^2})T.
\end{equation}
We will obtain a solution of the dmKPESCS (\ref{217}) as $V=q$,
$a_i=-1$ and $p_i=-q$, $i=1,\cdots,N$ with $q$ satisfies
(\ref{314}).

\section{Auto-B\"{a}cklund transformations for cdmKPH and for dmKPHSCS} \setcounter{equation}{0} \hskip\parindent
\begin{theorem}({\bf{Auto-B\"{a}cklund transformation for cdmKPH}})\\
Let ${\mathcal{L}}^n$ of (\ref{212}) satisfy the cdmKPH (\ref{214}).
$\phi$ is a function of $T$ satisfying
\begin{subequations}
\label{41}
\begin{equation}
\label{411}\phi_{T_k}={\mathcal{Q}}_k|_{p=\phi_X},\ \ \ \
k\in\mathbb{N},
\end{equation}
\begin{equation}
\label{412} {\mathcal{Q}}_n|_{p=\phi_X}-\s
(\frac{a_i}{\phi_X-p_i}+\frac{a_i}{p_i})=0.
\end{equation}
\end{subequations}
Define
$$A_1[\phi]:{\mathcal{L}}^n\mapsto {\mathcal{\tilde{L}}}^n=e^{-ad\phi}({\mathcal{L}}^n),$$
then $${\mathcal{\tilde{L}}}^n={\mathcal{\tilde{Q}}}_n-\s
(\frac{\tilde{a}_i}{\tilde{p}_i}+\frac{\tilde{a}_i}{p-\tilde{p}_i}),\
\ \ {\mathcal{\tilde{Q}}}_n=({\mathcal{\tilde{L}}}^n)_{\geq 1},$$
where $\tilde{a}_i=a_i$, $\tilde{p}_i=p_i-\phi_X$ will also satisfy
the cdmKPH (\ref{214}).
\end{theorem}
{\bf{Proof}}:
\begin{equation}
\begin{array}{lll}
({\mathcal{\tilde{L}}}^n)_{\leq
0}&=&[e^{-ad\phi}({\mathcal{L}}^n)]_{\leq
0}=e^{-ad\phi}(({\mathcal{L}}^n)_{\leq 0})+[e^{-ad\phi}(({\mathcal{L}}^n)_{\geq 1})]_0\\
&=&e^{-ad\phi}[-\s
(\frac{a_i}{p-p_i}+\frac{a_i}{p_i})]+({\mathcal{L}}^n)_{\geq
1}|_{p=\phi_X}\\
&=&-\s
(\frac{a_i}{p-(p_i-\phi_X)}+\frac{a_i}{p_i})+{\mathcal{Q}}_n|_{p=\phi_X}\\
&=&-\s\frac{a_i}{p-\tilde{p}_i}+\s\frac{a_i}{\phi_X-p_i}\\
&=&-\s(\frac{\tilde{a}_i}{\tilde{p}_i}+\frac{\tilde{a}_i}{p-\tilde{p}_i}).
\end{array}
\end{equation}
Furthermore, we have
\begin{equation}
\label{42}
\begin{array}{lll}
\tilde{{\mathcal{Q}}}_k&=&[({\mathcal{\tilde{L}}}^n)^{\frac{k}{n}}]_{\geq
1}=\{[e^{-ad\phi}({\mathcal{L}}^n)]^{\frac{k}{n}}\}_{\geq
1}=[e^{-ad\phi}(({\mathcal{L}}^n)^{\frac{k}{n}})]_{\geq 1}\\
&=&e^{-ad\phi}({\mathcal{Q}}_k)-{\mathcal{Q}}_k|_{p=\phi_X}=\sum_{l=0}^{\infty}\frac{1}{l!}\phi_X^l\p_{p}^l({\mathcal{Q}}_k)-{\mathcal{Q}}_k|_{p=\phi_X},
\end{array}
\end{equation}
\begin{equation}
\label{43}
(\tilde{{\mathcal{Q}}}_k)_p=[e^{-ad\phi}({\mathcal{Q}}_{k})]_p=e^{-ad\phi}({\mathcal{Q}}_{k,p})=\sum_{l=0}^{\infty}\frac{1}{l!}\phi_X^l\p_p^l({\mathcal{Q}}_{k,p}).
\end{equation}
So
\begin{subequations}
\label{44}
\begin{equation}
\label{441}
\tilde{{\mathcal{Q}}}_k|_{p=\tilde{p}_i}=\sum_{l=0}^{\infty}\frac{1}{l!}\phi_X^l\p_{p}^l({\mathcal{Q}}_k)|_{p=\tilde{p}_i}-{\mathcal{Q}}_k|_{p=\phi_X}={\mathcal{Q}}_k|_{p=p_i}-{\mathcal{Q}}_k|_{p=\phi_X},
\end{equation}
\begin{equation}
\label{442}
(\tilde{{\mathcal{Q}}}_k)_p|_{p=\tilde{p}_i}=\sum_{l=0}^{\infty}\frac{1}{l!}\phi_X^l\p_p^l({\mathcal{Q}}_{k,p})|_{p=\tilde{p}_i}={\mathcal{Q}}_{k,p}|_{p=p_i}.
\end{equation}
\end{subequations}
From Lemma 4.1 and (\ref{44}), we can see that
\begin{subequations}
\label{45}
\begin{equation}
\label{451}
(\tilde{{\mathcal{L}}}^n)_{T_k}-\{{\mathcal{\tilde{Q}}}_k,\tilde{{\mathcal{L}}}^n\}=e^{-ad\phi}[({\mathcal{L}}^n)_{T_k}-
\{{\mathcal{Q}}_k,{\mathcal{L}}^n\}]-\{\phi_{T_k}-{\mathcal{Q}}_k|_{p=\phi_X},\tilde{{\mathcal{L}}}^n\}=0,
\end{equation}
\begin{equation}
\label{452}
\tilde{a}_{i,T_k}=a_{i,T_k}=(a_i{\mathcal{Q}}_{k,p}|_{p=p_i})_X=(\tilde{a}_i(\tilde{{\mathcal{Q}}}_k)_p|_{p=\tilde{p}_i})_X,\
\ \ \ \ \ \ \ \ \ \ \ \ \ \ \ \ \ \ \ \ \ \ \ \ \ \ \ \ \
\end{equation}
\begin{equation}
\label{453}
\tilde{p}_{i,T_k}=p_{i,T_k}-(\phi_X)_{T_k}=({{\mathcal{Q}}}_k|_{p=p_i}-\phi_{T_k})_X=({{\mathcal{Q}}}_k|_{p=p_i}-{{\mathcal{Q}}}_k|_{p=\phi_X})_X=({\mathcal{\tilde{Q}}}_k|_{p=\tilde{p}_i})_X,
\end{equation}
\end{subequations}
i.e., $\tilde{{\mathcal{L}}}^n$, $\tilde{a}_i$, $\tilde{p}_i$ also
satisfy the cdmKPH
(\ref{214}).\\
This completes the proof.\\

\begin{theorem}({\bf{Auto-B\"{a}cklund transformation for dmKPHSCS}})\\
Let ${\mathcal{L}}^n$ of (\ref{212}) satisfy the dmKPHSCS
(\ref{215}) and $p$ satisfy the corresponding conservation equations
(\ref{216}). $\phi$ is a function of $T$ satisfying
\begin{subequations}
\label{46}
\begin{equation}
\label{461} \phi_{T_k}={\mathcal{Q}}_k|_{p=\phi_X},
\end{equation}
\begin{equation}
\label{462} \phi_{T_n}={\mathcal{Q}}_n|_{p=\phi_X}-\s(
\frac{a_i}{p_i}+\frac{a_i}{\phi_X-p_i}).
\end{equation}
\end{subequations}
Define
\begin{equation}
\label{463} A_2[\phi]:{\mathcal{L}}^n\mapsto
{\mathcal{\tilde{L}}}^n=[e^{-ad\phi}({\mathcal{L}}^n)]_{\geq 1}-\s
(\frac{\tilde{a}_i}{\tilde{p}_i}+\frac{\tilde{a}_i}{p-\tilde{p}_i}),
\end{equation}
where $\tilde{a}_i=a_i$, $\tilde{p}_i=p_i-\phi_X$, then
${\mathcal{\tilde{L}}}^n$ will also satisfy the dmKPHSCS
(\ref{215}).
\end{theorem}
{\bf{Proof}}: Define $\tilde{p}=p-\phi_X$. For $k<n$,
\begin{equation}
\label{47}
\begin{array}{lll}
\tilde{{\mathcal{Q}}}_k&=&[({\mathcal{\tilde{L}}}^n)^{\frac{k}{n}}]_{\geq
1}=[(({\mathcal{\tilde{L}}}^n)_{\geq 1})^{\frac{k}{n}}]_{\geq 1}=
\{[(e^{-ad\phi}({\mathcal{L}}^n))_{\geq 1}]^{\frac{k}{n}}\}_{\geq
1}=[(e^{-ad\phi}({\mathcal{L}}^n))^{\frac{k}{n}}]_{\geq 1}\\
&=&[e^{-ad\phi}(({\mathcal{L}}^n)^{\frac{k}{n}})]_{\geq
1}=e^{-ad\phi}({\mathcal{Q}}_k)-{\mathcal{Q}}_k|_{p=\phi_X}.
\end{array}
\end{equation}
So as proved in Theorem 4.1, $\tilde{a}_i$, $\tilde{p}_i$,
$i=1,\cdots,N$ will satisfy (\ref{2152}) and (\ref{2153})
respectively w.r.t. ${\mathcal{\tilde{L}}}^n$. Similarly we can
prove $\tilde{p}$ satisfies (\ref{2161}) w.r.t.
${\mathcal{\tilde{L}}}^n$.
\begin{equation}
\begin{array}{lll}
\tilde{p}_{T_n}&=&p_{T_n}-\phi_{X,T_n}=[{\mathcal{Q}}_n(p)-\s(\frac{a_i}{p_i}+\frac{a_i}{p-p_i})]_X-[{\mathcal{Q}}_n|_{p=\phi_X}-\s(\frac{a_i}{p_i}+\frac{a_i}{\phi_X-p_i})]_X\\
&=&({\mathcal{Q}}_n(p)-{\mathcal{Q}}_n|_{p=\phi_X})_X-\s(\frac{a_i}{p-p_i}-\frac{a_i}{\phi_X-p_i})_X\\
&=&[\tilde{{\mathcal{Q}}}_n(p)|_{p=\tilde{p}}]_X-\s(\frac{\tilde{a}_i}{\tilde{p}-\tilde{p}_i}+\frac{\tilde{a}_i}{\tilde{p}_i})_X,
\end{array}
\end{equation}
so $\tilde{p}$ satisfies (\ref{2162}) w.r.t.
${\mathcal{\tilde{L}}}^n$.\\
This completes the proof.\\
\\
{\bf{Example 2}:(Auto-B\"{a}cklund transformation for
dmKPESCS)}\\
When $n=3$, $A_2[\phi]$ will offers us an auto-B\"{a}cklund
transformation for the dmKPESCS
(\ref{217}). From (\ref{463}), $\tilde{V}=V+\phi_X$.\\
{\bf{(1)}} If we take a simple solution of the dmKPESCS (\ref{217})
as $V=0$, $a_i=0$, $p_i=1$, $i=1,\cdots,N$, then $\phi(X,Y,T)$
correspondingly also has to satisfy (\ref{3111}). Similarly like the
case in {\bf{Example 1} (1)}, we will obtain another degenerate
solution of the dmKPESCS (\ref{217}) as $\tilde{V}=q$,
$\tilde{a}_i=a_i=0$, $\tilde{p}_i=p_i-\phi_X=1-q$,
$i=1,\cdots,N$ with $q$ satisfying (\ref{313}).\\
{\bf{(2)}} If we choose $V=0$, $a_i=p_i=1$, $i=1,\cdots,N$, then via
$A_2[\phi]$, we can obtain a solution of the dmKPESCS (\ref{217}) as
$\tilde{V}=q$, $\tilde{a}_i=1$ and $\tilde{p}_i=1-q$, $i=1,\cdots,N$
with $q$ satisfies the following hodograph equation
\begin{equation}
X+2qY+[3q^2+\frac{N}{(q-1)^2}]=F(q),
\end{equation}
where $F(q)$ is an arbitrary function of $q$.

\section*{Acknowledgment}\hskip\parindent
This work was supported by the Chinese Basic Research Project
"Nonlinear Science".

\hskip\parindent

\end{document}